\documentstyle[aps,pra,epsfig]{revtex}

\begin{document}
\draft

\twocolumn[\hsize\textwidth\columnwidth
\hsize\csname @twocolumnfalse\endcsname

\title{The polymer theta-point as a knot delocalisation transition}
\author{Enzo Orlandini$^1$, Attilio L. Stella $^{1,2}$, and Carlo
Vanderzande$^{3,4}$}
\address{
$^1$ INFM-Dipartimento di Fisica, Universit\`{a} di Padova,
35131 Padova, Italy\\
$^2$ Sezione INFN, Universit\`{a} di Padova, 35131 Padova, Italy\\
$^3$ Departement WNI, Limburgs Universitair Centrum, Universitaire
Campus, 3590 Diepenbeek, Belgium\\
$^4$ Instituut voor Theoretische Fysica, Katholieke Universiteit Leuven, Celestijnenlaan 200D, 3001
Heverlee, Belgium}
\date{\today}
\maketitle

\begin{abstract}
We study numerically the tightness of prime flat knots in a model of
self-attracting
polymers with excluded volume. We find that these knots are localised in the
high temperature swollen regime, but become delocalised in the
low temperature globular phase. Precisely at the collapse
transition, the knots are weakly localised.
Some of our results can be interpreted in terms of the
theory of polymer networks, which allows to
conjecture exact exponents for the knot length probability distributions.
\end{abstract}
\hspace{.2in}
]

%\begin{multicols}{2} \narrowtext
The presence of knots in single ring macromolecules,
or of chain entanglement in polymer melts, has
fundamental consequences at the physical, chemical
and biological level. For example,
the dynamics of a ring polymer in a gel
depends crucially on its topology \cite{ORD}. The replication
and the transcription of circular DNA are controlled by
enzymes affecting the topology, the topoisomerases\cite{DNAboek}. 
Knots have even been
identified in some proteins in their native state \cite{Nat}.
It is easy to imagine the importance
that the degree of localisation of knots can have.
In the folding of a protein it must make quite a difference
whether the knot is tight and localised within a restricted
region of the backbone, or not.
A loose knot poses less
restrictions to the exploration of configuration space
in the search for the native state. Similarly, one
could expect that the degree of knot localisation
can strongly influence the function of topoisomerases.

These examples all concern heteropolymers
in nonequilibrium situations. Here we investigate
the interplay between topology and temperature
in a simpler context, by studying
the {\it equilibrium} properties of
(prime) knots in {\it homopolymers} when these are cooled
below their theta temperature. Under such conditions we find that
knots definitely loose their usual property of being
localized within restricted portions of the chains.  

It is very difficult to include topological
constraints within a statistical mechanical 
description of a polymer, since they imply a global control
of its conformations \cite{Vilgis}.
So far, most work has concentrated on the probability
of occurence of knots \cite{Prob} and
much less has been done on the physically relevant
problem of precisely quantifying the knot size.
Besides attempts to measure knot size directly \cite{Katrich},
most studies have given only indirect, and often incomplete information on knot
localisation.
For a ring polymer in the infinite temperature, good solvent
regime, 
it was found numerically that
the presence of a prime knot leads to
a simple multiplication of the partition sum
with a factor proportional to the length $L$ of the macromolecule \cite{OTJW1}.
Moreover, for the relation between the radius of gyration
and $L$ (Eq.(\ref{f1})), evidence was given
that neither the amplitude $A$ \cite{OTJW} nor the critical
exponent $\nu$ depends on the topology \cite{OTJW,Quake}.
These results suggest that knots
are somehow localised within the chain. On the other hand, in a recent
study of the force-extension relation for knotted
polymers in good solvent, corrections to scaling are interpreted in
terms of a length scale indirectly measuring the knot size
\cite{Kardar2}. From this, it is concluded that knots are only weakly localised,
in a sense that will be defined below.
 
Even if it is computationally quite hard to 
directly quantify the length of a knot in three-dimensional
space, the same is not true for
flat knots which result from the projection of a closed curve
onto a plane. Such projections are well known from knot theory
where they are at the basis of the determination of
topological invariants \cite{Knotboek}. 
Experimentally flat knots can be realized by adsorbing polymers
like DNA \cite{Ads}, or even macroscopic chains \cite{Exp} 
on a plane. In the former case thermal fluctuations 
can allow the system to reach equilibrium in two dimensions.

A flat knot can be drawn as
a number of crossings connected with
arcs. At each crossing some monomers have to
detach from the plane. 
If the adsorption interaction is sufficiently strong, any
non-minimal number of crossings becomes very unfavourable energetically.
For example, for the trefoil knot, this minimal number
is three and there are six arcs. In Ref. \cite{Kardar}
it was shown that for a ring of length $L$ with excluded volume
and at infinite temperature, there is typically
one of the arcs whose length is of order $L$,
whereas the total length of all the
other arcs, $l$, is much smaller: $l \ll L$.
More precisely, $\langle l \rangle$, the average value of $l$, was found not
to diverge with $L$. In this case one says that the knot
is localised. If in contrast, $\langle l \rangle \sim
L^t$, one speaks of weak localisation ($0<t<1$)
or delocalisation ($t=1$) of the knot.

In the present Letter, we investigate the size
of flat knots when the temperature $T$ is lowered, or
equivalently, when the quality of the solvent gets
worse. Under these circumstances the polymer will undergo a
collapse transition from a coil to a globule shape
below a theta point temperature $T_\theta$\cite{Boek}.
Our main result is that in the collapsed phase
knots are delocalised. At the theta-point,
we find them to be weakly localised with $t=3/7$.

A model for flat knots can be 
defined \cite{GuitOr} on the square lattice whose set
of edges is extended with the diagonals of the squares.
The bonds of the polymer can visit each edge and each vertex of
this extended lattice at most once. A diagonal
can only be occupied if at the same time the other,
perpendicular diagonal within the same elementary square is
also occupied. Each pair of occupied diagonals
represents a crossing (not to be considered as a lattice vertex)
in the projection of the knot.
Thus, one has to further specify
which of the two diagonal bonds goes under the other one.
The model is simulated within the
grand canonical ensemble, where a fugacity $K$ is
assigned to each bond of the ring polymer, while the number
of crossings is constrained to the minimum consistent with
the topology. As usual, to
include the possibility of theta collapse,
we associate an attractive energy with each pair of lattice vertices
that is visited by non-consecutive bonds.
Fig. 1 shows a configuration
with the topology of a trefoil.

\begin{figure}[]
\centerline{\hbox{
\epsfysize=15\baselineskip
        \epsffile{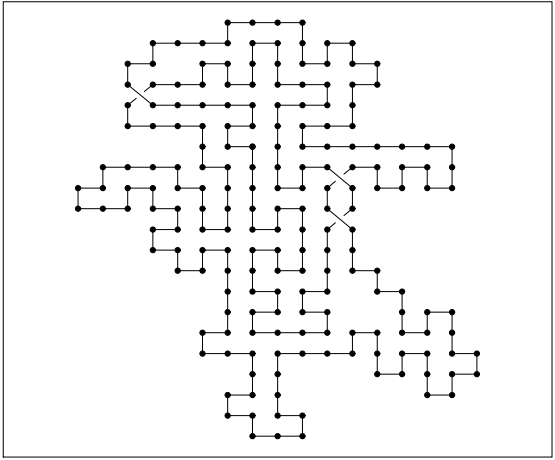}  }}
\vskip .3truecm
\caption [Figure]
	 {\protect\footnotesize
Typical configuration of a polymer with a trefoil knot in the low $T$
collapsed phase.
	 }
\label{fig1}
\end{figure}

In our Monte Carlo approach,
a Markov process in the configuration space
of the polymer is constructed by a combination of local
and non-local moves. These
are chosen as in Ref. \cite{GuitOr} and are such as to ensure invariance
of the polymer topology.
Averages at
fixed $T$ are then calculated
using a multiple
Markov chain (MMC) implementation in the fugacity $K$ \cite{MMC}.
This garantees an exhaustive sampling of configurations
also at rather low temperatures \cite{note3}.
First, we obtain precise estimates of $K_c(T)$,
the critical fugacity above which the grand canonical average $\langle L \rangle=\infty$. Since
for the case of a flat trefoil, there are always only
three pairs of diagonals occupied, we can expect
that $T_\theta$ is very close to the value
of an interacting self-avoiding ring model
without crossings.
%is $1/T_\theta=.65 \pm .03$\cite{FA}.
To verify this,
we investigated the average squared radius
of gyration $\langle R_g^2 \rangle_L$ \cite{note2} as a function of
$L$ for this case.
Fig. 2 reports our results for different $T$'s. One expects that asymptotically
\begin{eqnarray}
\langle R_g^2 \rangle_L \sim A L^{2 \nu}
\label{f1}
\end{eqnarray}
Our data clearly show the expected three regimes.
At high $T$'s, we are in a self-avoiding
walk regime with $\nu \simeq 3/4$. At low $T$'s
we determine an exponent $\nu=0.49 \pm .02$,
consistent with the value appropriate for a collapsed polymer,
$\nu=1/2$.
Finally, close to $1/T=0.67$
we find a $\nu$ in agreement
with that at the theta-point, i.e. $\nu=4/7$ \cite{DS}.
Hence, we estimate $1/T_{\theta}=0.67\pm.02$, fully
consistent with determinations for unknotted rings
\cite{FA}. 
We also conclude that the
exponent $\nu$ at the theta-point and in
the collapsed phase is not
modified by changing the topology from that of a flat
unknot to that of a flat trefoil.

\begin{figure}[]
\centerline{\hbox{
\epsfysize=17\baselineskip
        \epsffile{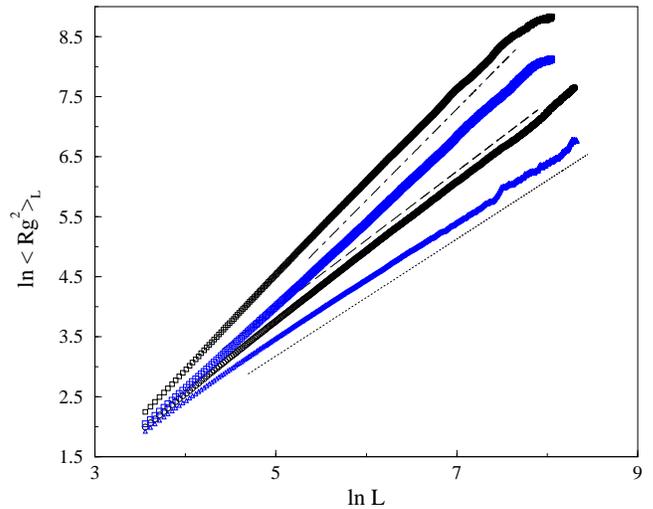}  }}
\caption [Figure]
	 {\protect\footnotesize
Log-Log plot of $\langle R_g^2 \rangle_L$ as a function
of $L$ for a ring with trefoil knot.
 From top to bottom : $1/T = 0, 0.50, 0.67 (\simeq 1/T_{\theta}), 0.8$.
The dot-dashed, the dashed, and the dotted lines have slopes $3/2$, $8/7$, and
$1.0$,
respectively.}
\label{fig2}
\end{figure}

In order to characterise the tightness of the trefoil knot,
we consider all its six arcs and determine the statistics
of their lengths $l_1 \leq l_2 \leq \ldots \leq l_6$.
Clearly the largest arc
must always have a length proportional to
$L$. In Fig. 3, we plot the average length of the second largest
arc, i.e. $\langle l_5 \rangle_L$, as a
function of $L$.
For
high $T$'s, we see that
$\langle l_5 \rangle_L$ saturates at large
enough $L$. Moreover, all the other arc lengths remain much smaller,
typically only a few bonds.
We conclude that in the
swollen regime the knot is localised.
The quantity $\langle l \rangle_L/L=\langle(\sum_{i=1}^{5} l_i)\rangle_L/L$ 
approaches zero when $L \to \infty$.

At $T_\theta$, we find instead
that $\langle l_5 \rangle_L$ grows
as $L^t$, with
$t = 0.44 \pm 0.02$. All other lengths remain
again very small. Thus, the typical shape of the
polymer appears to be that of a figure eight, as found also
at $T=\infty$ \cite{Kardar}.
We also conclude that the knot is weakly localised.

Finally, and most interestingly, we find that
in the collapsed phase, and for $L$-values that
are sufficiently large, $\langle l_5 \rangle_L \sim L$,
implying a {\it delocalisation} of the knot. In this case, there is
 ample evidence that also the average length of smaller arcs,
like $\langle l_4 \rangle_L$, starts to grow proportional to $L$ for
still longer polymer lengths. We suspect that
in sufficiently long polymers all average arc lengths
will become extensive in $L$. Thus, a description
in terms of a figure eight breaks down in
the collapsed phase.

\begin{figure}[]
\centerline{\hbox{
\epsfysize=17\baselineskip
        \epsffile{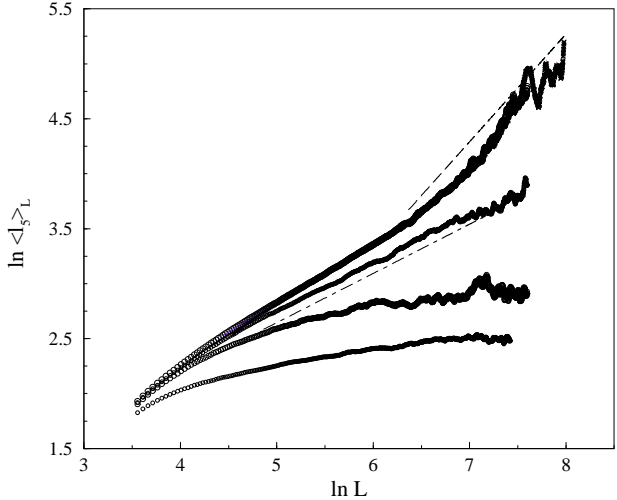}  }}
\caption [Figure]
	 {\protect\footnotesize
Log-Log plot of $\langle l_5 \rangle_L$  as a function
of $L$ for a trefoil knotted polymer.
  From bottom to top : $1/T = 0, 0.50, 0.67, 0.8$.
The dot-dashed line has a slope $0.44$ whereas
the dashed line has slope 1.0.}
\label{fig3}
\end{figure}
It is possible to associate a polymer network ${\cal G}$ with
$N$ segments \cite{Dupl}
to each flat knot configuration with the same number
of non-microscopic arc lengths \cite{Kardar}. Interest in polymer networks was revived recently by
a succesfull development concerning DNA denaturation \cite{Kafri}.
The same type of approach was subsequently applied in \cite{Kardar}
where it was found that flat knots with excluded volume
at $T=\infty$ are always localised. Our data for $T>T_\theta$ support this
conclusion, and moreover convincingly show that localisation
is present throughout the whole high temperature phase (Fig.3).

On the basis of a network description, this time with {\it interacting}
polymer segments, it is possible to gain further insight into some
of our results. Let ${\cal G}$ be such a network with
$n_k$ vertices of degree $k$ connected by
$N$ arcs of total length $L$.
The partition sum $Z_{{\cal G}}(l_1,\ldots,l_N)$
of the network scales as \cite{Dupl,Sch}
\begin{eqnarray}
Z_{{\cal G}}(l_1,\ldots,l_N)= K_c^{-L}~ l_N^{\gamma_{{\cal G}}-1}
F_{{\cal G}}\left( \frac{l_1}{l_N},\ldots,\frac{l_{N-1}}{l_N}\right)
\label{f2}
\end{eqnarray}
where $F_{{\cal G}}$ is a scaling function and
$\gamma_{{\cal G}}=1 - \nu d {\cal L} + \sum_k n_k \sigma_k$.
Here $d$ is the dimension of space and ${\cal L}$ is the number
of independent loops in the network. The lengths of the network segments
, corresponding to macroscopic knot arcs,
are $l_1 \leq ...\leq l_N$. Finally,
the exponents $\sigma_k$ are anomalous dimensions
associated to the $k$-leg vertices of the field theory
describing the polymer in the continuum limit \cite{Dupl}. In
two dimensions,
Coulomb gas methods allow an exact determination of
the $\sigma_k$'s.
In particular, at $T=\infty$,
where polymers are described by the $n \to 0$ limit of a critical
$O(n)$-model, one has $\sigma_k=(2-k)(9k+2)/64$ \cite{Dupl}. On the other hand,
the theta-point
is described by the critical low temperature phase of the
$O(n=1)$-model, for which $\sigma_k=(2-k)(2k+1)/42$ \cite{DS}.
Finally,
following \cite{DuplLT}, we assume that
the properties of the low $T$ regime can be
related to those of dense polymers, which are described
by the low $T$ phase of the $O(n=0)$-model.
For this case special care has to be taken,
and one ends up with a scaling form for $Z_{{\cal G}}$ which
is slightly different from Eq.(\ref{f2}) \cite{Dupldense}.
Yet, despite these differences, the network picture
can still be applied to flat knots with $\sigma_k=(4-k^2)/32$ \cite{Dupldense}.

When in a network one or more arcs become very short,
and hence the crossings on which they are incident approach each other
very closely,
the network should be replaced by another, contracted one with fewer segments and 
crossings, and a different $\gamma_{{\cal G}}$.
In this way, it can be understood that
even though any projection of a trefoil contains
six arcs, at a more coarse grained level, the typical contribution
% to
%Eq.(\ref{f2})
appearing in a numerical simulation
 can come from a
network with fewer crossings.
This is what happens, for example, at the theta-point, where
we found that
the flat trefoil looks
like a figure eight ($l_4 \sim O(1)$).
At $T_\theta$, where $\nu=4/7$, (\ref{f2}) predicts that the
partition function of a figure eight network scales as
\begin{eqnarray}
Z_{8} = K_c^{-L} (L-l)^{\gamma_8-1} F_8\left(\frac{l}{L-l}\right)
\label{f3}
\end{eqnarray}
with $\gamma_8=-12/7$.
For $l/L \to 0$, this partition sum should in its turn
reduce to that of a self-avoiding ring at the theta-point,
which is known to scale as $Z \sim K_c^{-L} L^{-\nu d}$. This simple analysis \cite{Kafri}
then teaches us that $F_8(x) \sim x^{-c}$ for $x \to 0$,
with $c=-(\gamma_8-1 + \nu d)=11/7 \approx 1.57$. Hence, for $l \ll L$
we predict that
\begin{eqnarray}
Z_{8} \sim K_c^{-L} (L-l)^{-\nu d}~l^{-c}
\label{f4}
\end{eqnarray}
In Fig. 4 (left) we present our data for $p(l_5)$,
the probability distribution of $l_5$, at $T=T_\theta$. From
(\ref{f4}) it follows that $p(l_5) \sim l_5^{-c}$.
A fit to this form leads to $c=1.63 \pm 0.08$,
consistent with the above prediction.
Finally, we get from (\ref{f4}) that $\langle l_5 \rangle_L \sim
L^{3/7}$, in good agreement with the numerical estimate
$\langle l_{5}\rangle_{L} \sim L^{.44\pm.02}$. We expect that $t=3/7$
is an exact result
which characterises the weak localisation of the knot at the theta-point.

At $T<T_\theta$, an analysis using the results of
\cite{Dupldense} can still be made for
a ring with the shape of a figure eight, and leads to
the prediction $c=11/8$. In Fig. 4 (right),
we also show our data for
$p(l_5)$ at $1/T=0.8>1/T_\theta$. There is
indeed an initial power law
decay with an exponent $1.34\pm.12$. But
in this case, $p(l_5)$ flattens for larger $l_5$-values
and it is this broadening which eventually leads to the delocalisation
of the knot. 
%\cite{note1}
The preasymptotic slope indicates
that for relatively small $l_5$, the weakly localised
figure eight network configurations are still dominating
the partition sum. When $\langle l_5 \rangle_L \sim L$
one can get no help from network scaling arguments
in determining $p(l_5)$. Indeed, such arguments are only
valid for $l_5 \ll L$ which is not the relevant range when $\langle l_5 \rangle_L \sim L$.

\begin{figure}[]
\centerline{\hbox{
\epsfysize=17\baselineskip
        \epsffile{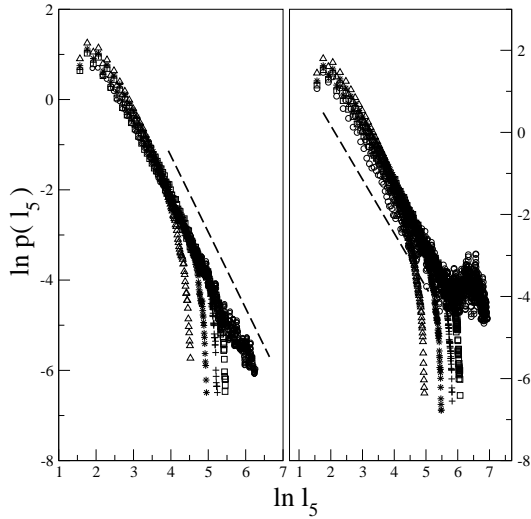}  }}
%\centerline{\epsfig{file=figure4.eps,height=6cm,width=7.5cm}}
\
\caption [Figure]
	 {\protect\footnotesize
Log-Log plot of
$p(l_{5})$. On the left (right), we show our data at $1/T_{\theta}\simeq .67
\ (1/T=.8)$. The dashed lines have slopes of respectively
$-1.63$ and $-1.34$. Different symbols refer to different
canonical averages with $L=200(\triangle),\ L=400(\ast),\ L=600(+),\ L=800(\Box)$ and $L=1400(\circ)$.
	 }
\label{fig4}
\end{figure}

We verify that also for other prime knots, such as the
$5_{1}$ and $7_{1}$ \cite{Knotboek},
  the typical
network configuration at $T=T_{\theta}$ is still the number eight
and delocalization occurs
for $T<T_\theta$.

In conclusion, we find that a flat prime knot in an adsorbed polymer
remains localised as long as $T>T_\theta$, becomes weakly localised at
$T_\theta$, and eventually delocalises at $T<T_\theta$. While delocalisation
is supported by strong numerical evidence, other results are fully consistent
with our predictions based on a polymer network approach. One may
speculate that the delocalisation found here could be a more general
phenomenon for the topology of interacting polymers.

Financial support from INFM-PAIS 01 and MIUR-COFIN 01 is gratefully
acknowledged.

\vskip -.4cm
%%%%%%%%%%%%%%%%%%%%%%%%%%%%%%%%%%%%%%%%%%%%%%%%%%%%%%%%%%

%\end{multicols}

\begin{references}
\bibitem{note3} Our MMC are based on
10 different $K$-values. For each one we sample $2\times 10^5$ independent configurations, corresponding
to $\sim 10^{11}$ local moves.
\bibitem{note2} $\langle . \rangle_L$ is a canonical average over
a subset of rings with length $\simeq L$.
\end{references}
\end{document}